\documentclass{aip-cp}

\usepackage[numbers]{natbib}
\usepackage{rotating}
\usepackage{graphicx}
\def\etal{{\it et al~}}
\begin{document}

\title{The Tungsten Project: Dielectronic Recombination Data For Collisional-Radiative Modelling In ITER}

\author[aff1]{S. P. Preval\corref{cor1}}
\author[aff1]{N. R. Badnell}
\author[aff1]{M. G. O'Mullane}

\affil[aff1]{Department of Physics, University of Strathclyde, Glasgow, G4 0NG, United Kingdom}
\corresp[cor1]{Corresponding author: simon.preval@strath.ac.uk}

\maketitle

\begin{abstract}
Tungsten is an important metal in nuclear fusion reactors. It will be used in the divertor component of ITER (Latin for `the 
way'). \textit{The Tungsten Project} aims to calculate partial and total DR rate coefficients for 
the isonuclear sequence of Tungsten. The calculated data will be made available as and when they are produced via the open access database OPEN-
ADAS in the standard \textit{adf09} and \textit{adf48} file formats. We present our progress thus far, detailing 
calculational methods, and showing comparisons with other available data. We conclude with plans for the future.

\end{abstract}

\section{INTRODUCTION}
Tungsten has been chosen as the plasma facing material that will compose the divertor in the upcoming experimental fusion reaction 
ITER. The reasons for this are threefold; tungsten has a high melting point, it is resistant to absorption of tritium,
and is also able to withstand with large power loads. In preparation for ITER, the Joint European Torus (JET) has been fitted with
a beryllium wall and tungsten divertor (\citet{matthews2011a}). This experimental setup is referred to as the ITER-like wall 
(ILW). Recent experiments with the ILW have shown reduced absorption of tritium as expected, but they have also shown that 
this configuration reduces the pedestal temperature in the tokamak from $\sim{1}$keV to $\sim{700}$eV (\citet{romanelli2013a}).

Being a plasma-facing component, tungsten will be sputtered into the divertor plasma, and will make its way into the main body plasma. This can potentially cool the plasma, leading to 
disruption and quenching. In order to understand the effect this impurity has on the plasma, detailed collisional-radiative 
(CR) modelling is required. The major caveat to this, however, is the provision of detailed, partial and total dielectronic 
recombination (DR) rate coefficient data.

\begin{figure}
\centerline{%
\includegraphics[width=120mm]{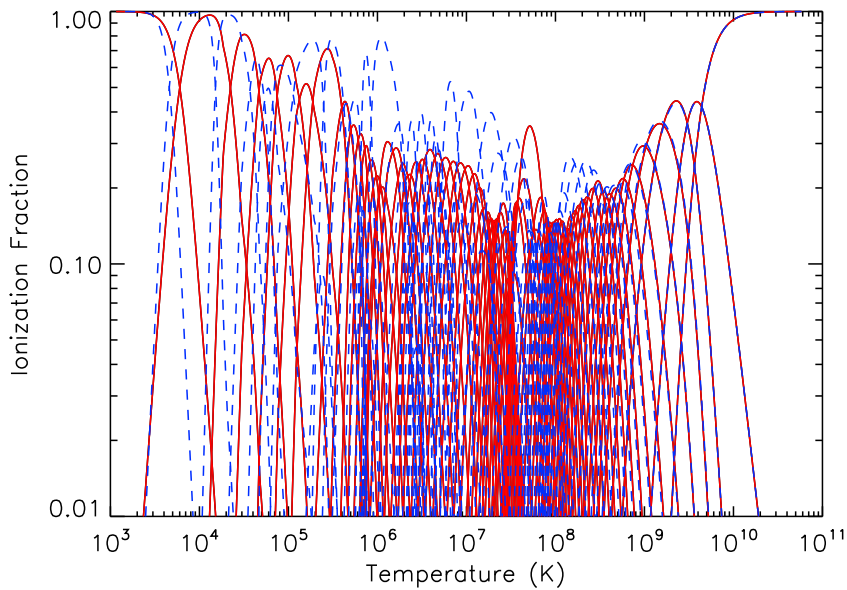}}
\caption{Comparison of ionization balances calculated using the recombination rate coefficients of \protect\citet{putterich2008a} (red 
solid curve) and \protect\citet{foster2008a} (blue dashed curve). Both curves use the ionization rate coefficients of \protect\citet{loch2005a}.}
\label{fig:ionbalputfos}
\end{figure}

This requirement has spurred multiple efforts to calculate such data. \citet{chung2005a} calculated DR rate coefficient data 
for the isonuclear sequence of tungsten using FLYCHK, which uses the average-atom method (\citet{zhao1997a}). \citet{putterich2008a}
used ADPAK (\citet{post1977a,post1995a}), which is based on the average-atom method also. In order to improve agreement with 
observation, the authors scaled the ADPAK rates for certain W stages. Finally, \citet{foster2008a} covered the isonuclear sequence using the Burgess 
General Formula (\citet{burgess1965a}). In both the P\"{u}tterich \etal and Foster isonuclear data, radiative recombination data was 
calculated using scaled hydrogenic values. In Figure \ref{fig:ionbalputfos} we have plotted the steady state ionization 
balance for tungsten using ionization rate coefficients from \citet{loch2005a}, and the recombination rate coefficients of 
P\"{u}tterich and Foster. It can be seen that large discrepancies exist for both the peak abundance temperatures, and the 
individual ionization fractions. 

In addition to isonuclear sequence work, several groups have also calculated detailed DR rate coefficients for individual 
ions. While not a complete list, a few examples are that of \citet{wu2015a}, who use the Flexible Atomic Code (\citet{gu2003a}) to 
calculate level-resolved DR rate coefficients for W$^{37+}$ -- W$^{46+}$, \citet{kwon2016a} who used {\sc hullac} 
(\citet{barshalom2001a}) to calculate W$^{45+}$, and \citet{peleg1998a} who also used {\sc hullac} to calculate W$^{56+}$. Again 
using {\sc hullac}, \citet{usafronova2009b}, and \citet{behar1999a} considered DR of W$^{63+}$. In addition, \citet{preval2016a} have 
used the {\sc autostructure} code (\citet{badnell1986a,badnell1997a,badnell2011a}) to calculate DR rate coefficients
for W$^{56+}$ -- W$^{73+}$. {\sc autostructure} has been experimentally verified in previous works such as \citet{savin2005a}.

As final-state-resolved partial DR rate coefficient data is only available for a handful of ions, we have launched 
\textit{The Tungsten Project}. The project will aim to calculate DR rate coefficients to as high an accuracy as possible for 
the entire isonuclear sequence of tungsten. In these proceedings, we will describe \textit{The Tungsten Project}, detailing 
calculational methods, and the data that will be generated. We conclude with a few remarks and the future work we plan to do.

\section{The Tungsten Project - Calculations}
\begin{figure}
\centerline{%
\includegraphics[width=120mm]{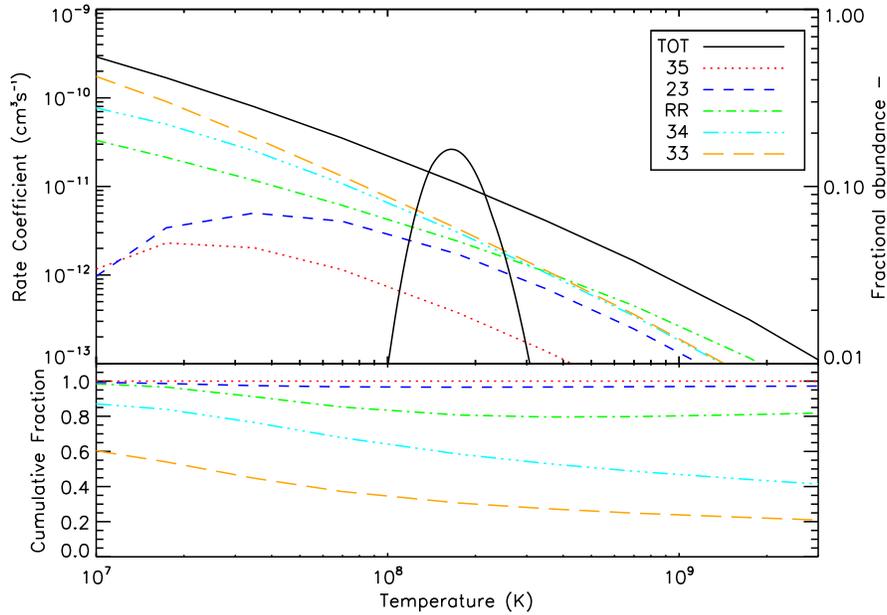}}
\caption{Plot of total DR and RR rate coefficients for 15-like W. The fractional abundance for 15-like is also shown (solid 
black parabola), which is calculated using the recombination data of \protect\citet{putterich2008a}, and the ionisation rate coefficients of \protect\citet{loch2005a}. 
The bottom plot shows the cumulative fraction for each core excitation to the total recombination rate coefficient.}
\label{fig:rates}
\end{figure}

As a technical point, different ionisation stages of tungsten will be referred to by the number of valence electrons present 
in said ion. For example, H-like (one electron) becomes 01-like, and Si-like (14 electrons) becomes 14-like. A detailed 
description of the calculational method can be found in \citet{preval2016a}, however, we provide a brief summary here. All 
calculations were carried out using the distorted wave code {\sc autostructure} (\citet{badnell1986a,badnell1997a,badnell2011a}). 
{\sc autostructure} is able to calculate energy levels, oscillator strengths,  photoionization cross sections, and many other 
quantities. These quantites can be calculated with level resolution (intermediate coupling IC), term resolution (LS coupling), or 
configuration resolution (configuration average CA) using semi-relativistic kappa-averaged wavefunctions. We label our DR 
rate coefficient calculations for each ion by the core excitation considered, which is described by the initial and final $n$ 
of the DR process ($n_{i}$ and $n_{f}$ respectively). As an example, a core excitation with $n_{i}=3$ and $n_{f}=4$ is 
labeled as 3--4. To decide which core excitations should be calculated in IC for a particular ion, we first calculate them in 
CA and compare their contribution to the overall total. If it exceeds 5\%, we perform the calculation in IC. In addition 
to calculating DR rate coefficients, we also calculate radiative recombination (RR) rate coefficients, which are important for 00-like to 18-like.
In Figure \ref{fig:rates} we give an example of calculated IC DR/RR rate 
coefficients for 15-like tungsten. The figure also shows the cumulative fraction, which shows the relative contribution of each core excitation and 
the RR rate coefficient to the total recombination rate coefficient. The largest contribution is given first, followed by the largest plus the second 
largest, and so on.

\subsection{DR}
The Rydberg electron DR is calculated for each explicit principal quantum number $n$ up to $n=25$, after which it is then 
calculated for $n$ on a logarithmic scale up to $n=999$. Interpolation is then used to calculate DR for intermediate values of $n$.
Angular momenta $\ell$ for each $n$ are calculated up to a maximum value such that the DR rate coefficients are numerically 
converged to $<1$\% over the entire ADAS temperature range $z^{2}(10-10^{7})$K, where $z$ is the residual charge. For $n$ 
values greater than $(\frac{1}{4}\ell^2 + 15)$ {\sc autostructure} approximates a bound wavefunction $P_{n\ell}$ by a zero 
energy continuum wavefunction $F_{k\ell}(r)$ using the result (\citet{badnell1989a})
\begin{equation}
\lim_{n\rightarrow\infty}\left(\frac{\pi\nu_{n\ell}^3}{2z^2}\right)^{\frac{1}{2}}P_{n\ell}(r)=F_{k\ell}(r)|_{k=0},
\end{equation}
where $k^{2}=E$, $\nu_{nl}=n-\mu_{\ell}$, and $\mu_{\ell}$ is the quantum defect. This approach allows the Hamiltonian to be 
diagonalized for all $n$ and to allow for the opening up of new Auger channels at high $n$.

The configurations for the N-electron target consist of all single excitations from each subshell. The N+1 configurations are 
simply the N-electron configurations with an additional electron. In Table \ref{table:conf} we include an example of the 
configurations used for 15-like 3--3. Mixing configurations are also included by way of the ``one up-one down'' rule, as these 
configurations are typically close to each other in terms of energy, and hence mix strongly.

\begin{table*}
\caption{List of configurations included in IC calculation for 15-like 3--3. The left column gives the N-electron 
configurations, whilst the right gives the N+1-electron 
configurations. Configurations marked with * are included as mixing configurations.}
\begin{centering}
\begin{tabular}{ll}
\hline
N-electron & N+1-electron \\
\hline 
$3s^{2} 3p^{3}$     & $3s^{2} 3p^{4}$        \\
$3s^{2} 3p^{2} 3d$  & $3s^{2} 3p^{3} 3d$     \\
$3s 3p^{4}$         & $3s^{2} 3p^{2} 3d^{2}$ \\
$3s 3p^{3} 3d$      & $3s 3p^{5}$            \\
$*3p^{5}$           & $3s 3p^{4} 3d$         \\
$*3s^{2} 3p 3d^{2}$ & $3s 3p^{3} 3d^{2}$     \\
                    & $*3p^{6}$              \\
                    & $*3p^{5} 3d$           \\
                    & $*3s^{2} 3p 3d^{3}$    \\
\hline
\label{table:conf}
\end{tabular}
\end{centering}
\end{table*}

\subsection{RR}
As with DR, the Rydberg electron RR is calculated for each $n$ up to $n=25$, and then logarithmically up to $n=999$. $\ell$ 
are included relativistically up to $\ell=10$, after which a non-relativistic top-up up to $\ell=150$ is added. This is 
sufficient to converge bare-like RR to $<1$\% over the ADAS temperature range as given above. The 
N-electron configurations used for RR calculations resemble that of the $\Delta{n}=0$ core excitations for DR. All possible 
excitations of the outermost electron are included plus mixing. Again, the N+1 configurations are just the N-electron 
configurations plus an electron. The RR rate coefficients include multipolar radiation contributions up to E40 and M39.

\section{Comparisons}

\subsection{Ionization Balance}
\begin{figure}
\centerline{%
\includegraphics[width=120mm]{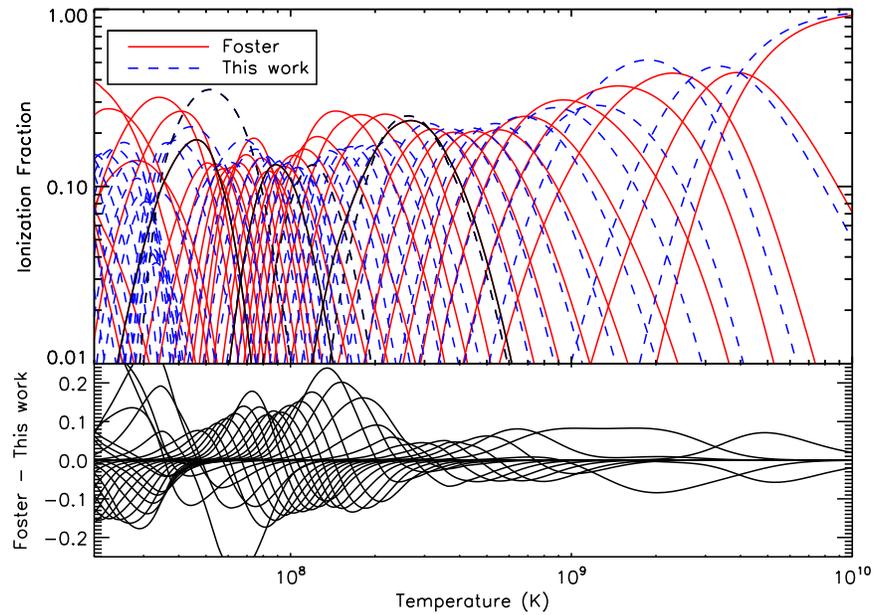}}
\caption{Comparison of ionization balances using recombination data calculated in this work (blue dashed curve),
and that of \protect\citet{foster2008a} (red solid curve) up to 18-like tungsten. Again, both curves use ionization rate 
coefficients from \protect\citet{loch2005a}. The black lines indicate (from right to left) the positions of 10-like and 18-like 
tungsten.}
\label{fig:ionbalprefos}
\end{figure}

In Figure \ref{fig:ionbalprefos} we compare the ionization balance calculated using the \citet{foster2008a} recombination data, 
and our present calculations. Both recombination data sets use the ionization rate coefficients of \citet{loch2005a}. For our 
ionization balance, we include data up to 18-like, after which we use \citet{putterich2008a}'s data to 74-like. We can see that the 
agreement between our ionization balance and \citet{foster2008a}'s is generally poor, but consistently so up to 10-like. This is because 
the contribution from DR is small compared to that of RR from 01-like ($\sim{99}$\%). This contribution from DR gradually 
increases until it becomes comparable to RR at 10-like, and goes on the exceed this as the residual charge of the ion 
decreases. Another factor causing the difference is the application of the J\"{u}ttner relativistic correction to the DR/RR rate 
coefficients for our calculation, but not that of Foster's.

\subsection{10-like tungsten}
\begin{figure}
\centerline{%
\includegraphics[width=120mm]{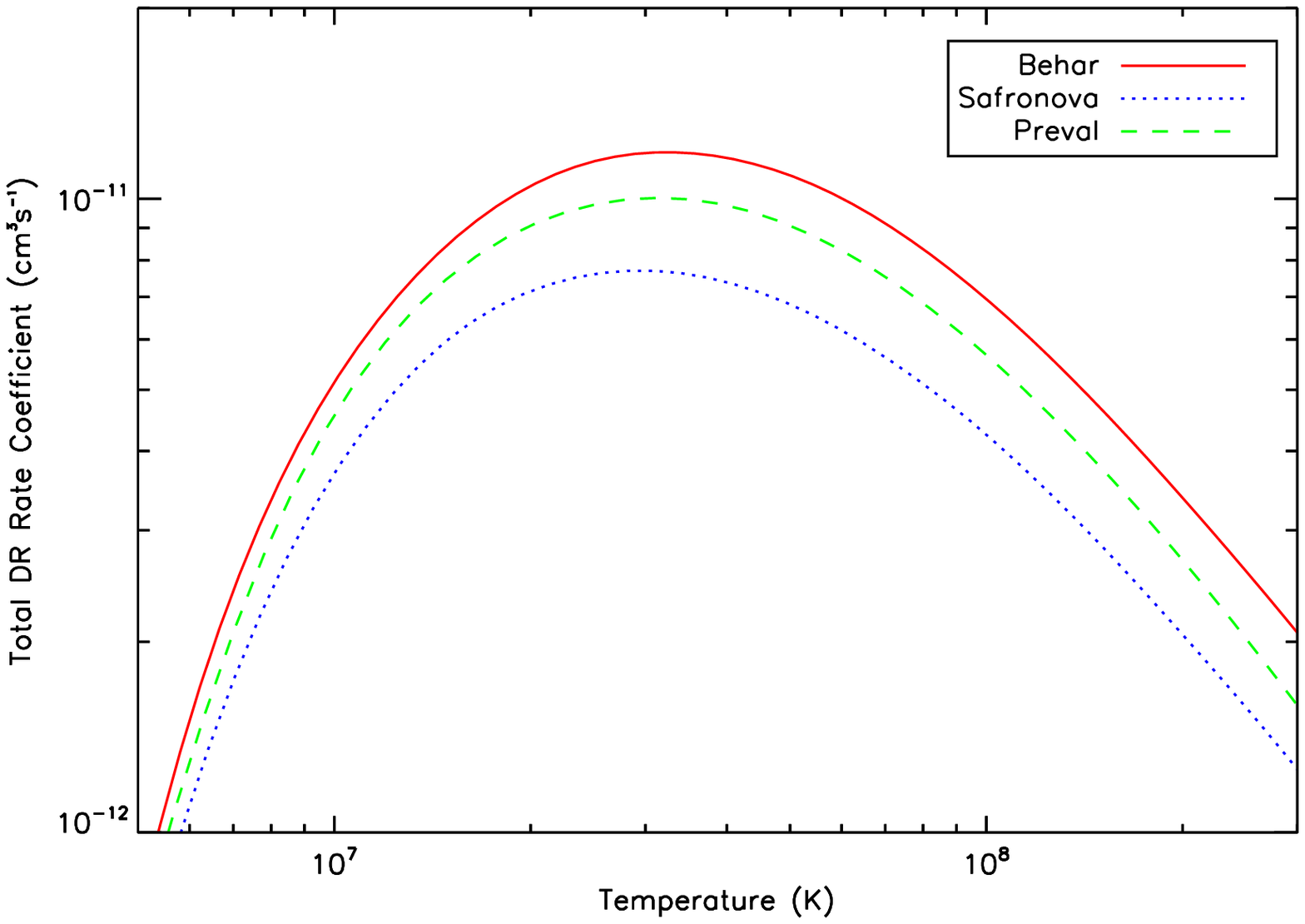}}
\caption{Comparison of total DR rate coefficients for 10-like tungsten calculated by \protect\citet{behar1999a} (red solid line), 
\protect\citet{usafronova2009b} (blue dashed line) and in this work (green dotted line).}
\label{fig:necomp}
\end{figure}

As mentioned in the Introduction, DR of 10-like tungsten has been considered by \citet{behar1999a} and \citet{usafronova2009b}. In 
Figure \ref{fig:necomp} we compare the total DR rate coefficients calculated in this work with the data calculated by Behar \etal 
and Safronova \etal. It can be seen that our rate coefficients sit somewhere between the two. Typically, 
the difference between our rate coefficients and Behar \etal's do not exceed 10\% at peak abundance temperature, whereas the 
difference is typically $>40$\% when comparing to Safronova \etal's data. These differences may arise due to the way the
higher-$n$ DR is calculated. In the case of Behar \etal's and Safronova \etal's data, radiative and Auger rates are calculated up 
to a nominal value of $n$ (typically 8--15), which are then extrapolated up to $n\sim{1000}$. This extrapolation method may miss additional Auger channels 
opening-up at higher-$n$, whereas our method described previously allows for their inclusion in the calculation. In the future, it will be prudent to assess the 
sensitivity of the extrapolation method to the final explicit $n$ calculated before extrapolation is used. 

\section{Future work and conclusions}
To date, we have completed calculations for 01-like to 37-like. Calculations 
are currently ongoing for 38-like -- 57-like. All data that will be generated will be hosted on the OPEN-ADAS 
website\footnote{https://www.open.adas.ac.uk} in the standard \textit{adf09} and \textit{adf48} formats. 
In these proceedings we have introduced \textit{The Tungsten Project}, and have described its aims, and 
the data that will be produced. We have used {\sc autostructure} to calculate partial DR/RR rate coefficient data for 
ionization stages 01-like -- 37-like tungsten. One paper has been published covering 01-like -- 18-like, and we are currently 
writing a second paper for 19-like -- 36-like. We are currently conducting calculations covering the $4d$ and $4f$ shells, and 
expect these to be completed in a timely fashion. We are aiming to calculate updated CR models incorporating 
the new data described here.

\section{ACKNOWLEDGMENTS}
This work was supported by the Engineering and Physical Sciences Research Council (EPSRC), Grant No. EP/1021803 to the 
University of Strathclyde.


\begin{thebibliography}{0}%
\makeatletter
\providecommand \@ifxundefined [1]{%
 \@ifx{#1\undefined}
}%
\providecommand \@ifnum [1]{%
 \ifnum #1\expandafter \@firstoftwo
 \else \expandafter \@secondoftwo
 \fi
}%
\providecommand \@ifx [1]{%
 \ifx #1\expandafter \@firstoftwo
 \else \expandafter \@secondoftwo
 \fi
}%
\providecommand \natexlab [1]{#1}%
\providecommand \enquote  [1]{``#1''}%
\providecommand \bibnamefont  [1]{#1}%
\providecommand \bibfnamefont [1]{#1}%
\providecommand \citenamefont [1]{#1}%
\providecommand \href@noop [0]{\@secondoftwo}%
\providecommand \href [0]{\begingroup \@sanitize@url \@href}%
\providecommand \@href[1]{\@@startlink{#1}\@@href}%
\providecommand \@@href[1]{\endgroup#1\@@endlink}%
\providecommand \@sanitize@url [0]{\catcode `\$12\catcode `\&12\catcode
  `\#12\catcode `\^12\catcode `\_12\catcode `\%12\relax}%
\providecommand \@@startlink[1]{}%
\providecommand \@@endlink[0]{}%
\providecommand \url  [0]{\begingroup\@sanitize@url \@url }%
\providecommand \@url [1]{\endgroup\@href {#1}{\urlprefix }}%
\providecommand \urlprefix  [0]{URL }%
\providecommand \Eprint [0]{\href }%
\providecommand \doibase [0]{http://dx.doi.org/}%
\providecommand \selectlanguage [0]{\@gobble}%
\providecommand \bibinfo  [0]{\@secondoftwo}%
\providecommand \bibfield  [0]{\@secondoftwo}%
\providecommand \translation [1]{[#1]}%
\providecommand \BibitemOpen [0]{}%
\providecommand \bibitemStop [0]{}%
\providecommand \bibitemNoStop [0]{.\EOS\space}%
\providecommand \EOS [0]{\spacefactor3000\relax}%
\providecommand \BibitemShut  [1]{\csname bibitem#1\endcsname}%
\let\auto@bib@innerbib\@empty
\end{thebibliography}%


\begin{thebibliography}{22}%
\makeatletter
\providecommand \@ifxundefined [1]{%
 \@ifx{#1\undefined}
}%
\providecommand \@ifnum [1]{%
 \ifnum #1\expandafter \@firstoftwo
 \else \expandafter \@secondoftwo
 \fi
}%
\providecommand \@ifx [1]{%
 \ifx #1\expandafter \@firstoftwo
 \else \expandafter \@secondoftwo
 \fi
}%
\providecommand \natexlab [1]{#1}%
\providecommand \enquote  [1]{``#1''}%
\providecommand \bibnamefont  [1]{#1}%
\providecommand \bibfnamefont [1]{#1}%
\providecommand \citenamefont [1]{#1}%
\providecommand \href@noop [0]{\@secondoftwo}%
\providecommand \href [0]{\begingroup \@sanitize@url \@href}%
\providecommand \@href[1]{\@@startlink{#1}\@@href}%
\providecommand \@@href[1]{\endgroup#1\@@endlink}%
\providecommand \@sanitize@url [0]{\catcode `\$12\catcode `\&12\catcode
  `\#12\catcode `\^12\catcode `\_12\catcode `\%12\relax}%
\providecommand \@@startlink[1]{}%
\providecommand \@@endlink[0]{}%
\providecommand \url  [0]{\begingroup\@sanitize@url \@url }%
\providecommand \@url [1]{\endgroup\@href {#1}{\urlprefix }}%
\providecommand \urlprefix  [0]{URL }%
\providecommand \Eprint [0]{\href }%
\providecommand \doibase [0]{http://dx.doi.org/}%
\providecommand \selectlanguage [0]{\@gobble}%
\providecommand \bibinfo  [0]{\@secondoftwo}%
\providecommand \bibfield  [0]{\@secondoftwo}%
\providecommand \translation [1]{[#1]}%
\providecommand \BibitemOpen [0]{}%
\providecommand \bibitemStop [0]{}%
\providecommand \bibitemNoStop [0]{.\EOS\space}%
\providecommand \EOS [0]{\spacefactor3000\relax}%
\providecommand \BibitemShut  [1]{\csname bibitem#1\endcsname}%
\let\auto@bib@innerbib\@empty
\bibitem [{\citenamefont {Matthews}\ \emph {et~al.}(2011)\citenamefont
  {Matthews}, \citenamefont {Beurskens}, \citenamefont {Brezinsek},
  \citenamefont {Groth}, \citenamefont {Joffrin}, \citenamefont {Loving},
  \citenamefont {Kear}, \citenamefont {Mayoral}, \citenamefont {Neu},
  \citenamefont {Prior}, \citenamefont {Riccardo}, \citenamefont {Rimini},
  \citenamefont {Rubel}, \citenamefont {Sips}, \citenamefont {Villedieu},
  \citenamefont {de~Vries}, \citenamefont {Watkins},\ and\ \citenamefont
  {Contributors}}]{matthews2011a}%
  \BibitemOpen
  \bibfield  {author} {\bibinfo {author} {\bibfnamefont {G.~F.}\ \bibnamefont
  {Matthews}}, \bibinfo {author} {\bibfnamefont {M.}~\bibnamefont {Beurskens}},
  \bibinfo {author} {\bibfnamefont {S.}~\bibnamefont {Brezinsek}}, \bibinfo
  {author} {\bibfnamefont {M.}~\bibnamefont {Groth}}, \bibinfo {author}
  {\bibfnamefont {E.}~\bibnamefont {Joffrin}}, \bibinfo {author} {\bibfnamefont
  {A.}~\bibnamefont {Loving}}, \bibinfo {author} {\bibfnamefont
  {M.}~\bibnamefont {Kear}}, \bibinfo {author} {\bibfnamefont {M.~L.}\
  \bibnamefont {Mayoral}}, \bibinfo {author} {\bibfnamefont {R.}~\bibnamefont
  {Neu}}, \bibinfo {author} {\bibfnamefont {P.}~\bibnamefont {Prior}}, \bibinfo
  {author} {\bibfnamefont {V.}~\bibnamefont {Riccardo}}, \bibinfo {author}
  {\bibfnamefont {F.}~\bibnamefont {Rimini}}, \bibinfo {author} {\bibfnamefont
  {M.}~\bibnamefont {Rubel}}, \bibinfo {author} {\bibfnamefont
  {G.}~\bibnamefont {Sips}}, \bibinfo {author} {\bibfnamefont {E.}~\bibnamefont
  {Villedieu}}, \bibinfo {author} {\bibfnamefont {P.}~\bibnamefont {de~Vries}},
  \bibinfo {author} {\bibfnamefont {M.~L.}\ \bibnamefont {Watkins}}, \ and\
  \bibinfo {author} {\bibfnamefont {E.-J.}\ \bibnamefont {Contributors}},\
  }\href {http://stacks.iop.org/1402-4896/2011/i=T145/a=014001} {\bibfield
  {journal} {\bibinfo  {journal} {Physica Scripta}\ }\textbf {\bibinfo {volume}
  {2011}},\  \bibinfo {pages} {014001} (\bibinfo {year} {2011})}\BibitemShut
  {NoStop}%
\bibitem [{\citenamefont {Romanelli}\ and\ \citenamefont
  {Contributors}(2013)}]{romanelli2013a}%
  \BibitemOpen
  \bibfield  {author} {\bibinfo {author} {\bibfnamefont {F.}~\bibnamefont
  {Romanelli}}\ and\ \bibinfo {author} {\bibfnamefont {J.~E. F. D.~A.}\
  \bibnamefont {Contributors}},\ }\href
  {http://stacks.iop.org/0029-5515/53/i=10/a=104002} {\bibfield  {journal}
  {\bibinfo  {journal} {Nuclear Fusion}\ }\textbf {\bibinfo {volume} {53}},\
   \bibinfo {pages} {104002} (\bibinfo {year} {2013})}\BibitemShut {NoStop}%
\bibitem [{\citenamefont {Chung}\ \emph {et~al.}(2005)\citenamefont {Chung},
  \citenamefont {Chen}, \citenamefont {Morgan}, \citenamefont {Ralchenko},\
  and\ \citenamefont {Lee}}]{chung2005a}%
  \BibitemOpen
  \bibfield  {author} {\bibinfo {author} {\bibfnamefont {H.~K.}\ \bibnamefont
  {Chung}}, \bibinfo {author} {\bibfnamefont {M.~H.}\ \bibnamefont {Chen}},
  \bibinfo {author} {\bibfnamefont {W.~L.}\ \bibnamefont {Morgan}}, \bibinfo
  {author} {\bibfnamefont {Y.}~\bibnamefont {Ralchenko}}, \ and\ \bibinfo
  {author} {\bibfnamefont {R.~W.}\ \bibnamefont {Lee}},\ }\href {\doibase
  http://dx.doi.org/10.1016/j.hedp.2005.07.001} {\bibfield  {journal} {\bibinfo
   {journal} {High Energy Density Physics}\ }\textbf {\bibinfo {volume} {1}},\
  \unskip\ \bibinfo {pages} {3} (\bibinfo {year} {2005})}\BibitemShut
  {NoStop}%
\bibitem [{\citenamefont {Zhao}\ and\ \citenamefont {Li}(1997)}]{zhao1997a}%
  \BibitemOpen
  \bibfield  {author} {\bibinfo {author} {\bibfnamefont {L.~B.}\ \bibnamefont
  {Zhao}}\ and\ \bibinfo {author} {\bibfnamefont {S.~C.}\ \bibnamefont {Li}},\
  }\href {\doibase 10.1103/PhysRevA.55.1039} {\bibfield  {journal} {\bibinfo
  {journal} {Phys. Rev. A}\ }\textbf {\bibinfo {volume} {55}},\ \unskip\
  \bibinfo {pages} {1039} (\bibinfo {year} {1997})}\BibitemShut {NoStop}%
\bibitem [{\citenamefont {P\"{u}tterich}\ \emph {et~al.}(2008)\citenamefont
  {P\"{u}tterich}, \citenamefont {Neu}, \citenamefont {Dux}, \citenamefont
  {Whiteford}, \citenamefont {O'Mullane},\ and\ \citenamefont {the ASDEX
  Upgrade~Team}}]{putterich2008a}%
  \BibitemOpen
  \bibfield  {author} {\bibinfo {author} {\bibfnamefont {T.}~\bibnamefont
  {P\"{u}tterich}}, \bibinfo {author} {\bibfnamefont {R.}~\bibnamefont {Neu}},
  \bibinfo {author} {\bibfnamefont {R.}~\bibnamefont {Dux}}, \bibinfo {author}
  {\bibfnamefont {A.~D.}\ \bibnamefont {Whiteford}}, \bibinfo {author}
  {\bibfnamefont {M.~G.}\ \bibnamefont {O'Mullane}}, \ and\ \bibinfo {author}
  {\bibnamefont {the ASDEX Upgrade~Team}},\ }\href
  {http://stacks.iop.org/0741-3335/50/i=8/a=085016} {\bibfield  {journal}
  {\bibinfo  {journal} {Plasma Physics and Controlled Fusion}\ }\textbf
  {\bibinfo {volume} {50}},\ \bibinfo {pages} {085016} (\bibinfo {year}
  {2008})}\BibitemShut {NoStop}%
\bibitem [{\citenamefont {Post}\ \emph {et~al.}(1977)\citenamefont {Post},
  \citenamefont {Jensen}, \citenamefont {Tarter}, \citenamefont {Grasberger},\
  and\ \citenamefont {Lokke}}]{post1977a}%
  \BibitemOpen
  \bibfield  {author} {\bibinfo {author} {\bibfnamefont {D.~E.}\ \bibnamefont
  {Post}}, \bibinfo {author} {\bibfnamefont {R.~V.}\ \bibnamefont {Jensen}},
  \bibinfo {author} {\bibfnamefont {C.~B.}\ \bibnamefont {Tarter}}, \bibinfo
  {author} {\bibfnamefont {W.~H.}\ \bibnamefont {Grasberger}}, \ and\ \bibinfo
  {author} {\bibfnamefont {W.~A.}\ \bibnamefont {Lokke}},\ }\href {\doibase
  http://dx.doi.org/10.1016/0092-640X(77)90026-2} {\bibfield  {journal}
  {\bibinfo  {journal} {Atomic Data and Nuclear Data Tables}\ }\textbf
  {\bibinfo {volume} {20}},\ \unskip\ \bibinfo {pages} {397} (\bibinfo
  {year} {1977})}\BibitemShut {NoStop}%
\bibitem [{\citenamefont {Post}\ \emph {et~al.}(1995)\citenamefont {Post},
  \citenamefont {Abdallah}, \citenamefont {Clark},\ and\ \citenamefont
  {Putvinskaya}}]{post1995a}%
  \BibitemOpen
  \bibfield  {author} {\bibinfo {author} {\bibfnamefont {D.}~\bibnamefont
  {Post}}, \bibinfo {author} {\bibfnamefont {J.}~\bibnamefont {Abdallah}},
  \bibinfo {author} {\bibfnamefont {R.~E.~H.}\ \bibnamefont {Clark}}, \ and\
  \bibinfo {author} {\bibfnamefont {N.}~\bibnamefont {Putvinskaya}},\ }\href
  {\doibase http://dx.doi.org/10.1063/1.871257} {\bibfield  {journal} {\bibinfo
   {journal} {Physics of Plasmas}\ }\textbf {\bibinfo {volume} {2}},\ \unskip\
  \bibinfo {pages} {2328} (\bibinfo {year} {1995})}\BibitemShut {NoStop}%
\bibitem [{\citenamefont {Foster}(2008)}]{foster2008a}%
  \BibitemOpen
  \bibfield  {author} {\bibinfo {author} {\bibfnamefont {A.~R.}\ \bibnamefont
  {Foster}},\ }\enquote {\bibinfo {title} {{On the Behaviour and Radiating
  Properties of Heavy Elements in Fusion Plasmas}},}\ \href@noop {} {Ph.D.
  thesis},\ \bibinfo  {school} {University of Strathclyde}, \bibinfo {address}
  {http://www.adas.ac.uk/theses/foster\_thesis.pdf} \bibinfo {year}
  {2008}\BibitemShut {NoStop}%
\bibitem [{\citenamefont {{Burgess}}(1965)}]{burgess1965a}%
  \BibitemOpen
  \bibfield  {author} {\bibinfo {author} {\bibfnamefont {A.}~\bibnamefont
  {{Burgess}}},\ }\href {\doibase 10.1086/148253} {\bibfield  {journal}
  {\bibinfo  {journal} {ApJ}\ }\textbf {\bibinfo {volume} {141}},\ \unskip\
  \bibinfo {pages} {1588} (\bibinfo {year} {1965})}\BibitemShut {NoStop}%
\bibitem [{\citenamefont {Loch}\ \emph {et~al.}(2005)\citenamefont {Loch},
  \citenamefont {Ludlow}, \citenamefont {Pindzola}, \citenamefont {Whiteford},\
  and\ \citenamefont {Griffin}}]{loch2005a}%
  \BibitemOpen
  \bibfield  {author} {\bibinfo {author} {\bibfnamefont {S.~D.}\ \bibnamefont
  {Loch}}, \bibinfo {author} {\bibfnamefont {J.~A.}\ \bibnamefont {Ludlow}},
  \bibinfo {author} {\bibfnamefont {M.~S.}\ \bibnamefont {Pindzola}}, \bibinfo
  {author} {\bibfnamefont {A.~D.}\ \bibnamefont {Whiteford}}, \ and\ \bibinfo
  {author} {\bibfnamefont {D.~C.}\ \bibnamefont {Griffin}},\ }\href {\doibase
  10.1103/PhysRevA.72.052716} {\bibfield  {journal} {\bibinfo  {journal} {Phys.
  Rev. A}\ }\textbf {\bibinfo {volume} {72}},\ \unskip\ \bibinfo {pages}
  {052716} (\bibinfo {year} {2005})}\BibitemShut {NoStop}%
\bibitem [{\citenamefont {Wu}\ \emph {et~al.}(2015)\citenamefont {Wu},
  \citenamefont {Fu}, \citenamefont {Ma}, \citenamefont {Li}, \citenamefont
  {Xie}, \citenamefont {Jiang},\ and\ \citenamefont {Dong}}]{wu2015a}%
  \BibitemOpen
  \bibfield  {author} {\bibinfo {author} {\bibfnamefont {Z.}~\bibnamefont
  {Wu}}, \bibinfo {author} {\bibfnamefont {Y.}~\bibnamefont {Fu}}, \bibinfo
  {author} {\bibfnamefont {X.}~\bibnamefont {Ma}}, \bibinfo {author}
  {\bibfnamefont {M.}~\bibnamefont {Li}}, \bibinfo {author} {\bibfnamefont
  {L.}~\bibnamefont {Xie}}, \bibinfo {author} {\bibfnamefont {J.}~\bibnamefont
  {Jiang}}, \ and\ \bibinfo {author} {\bibfnamefont {C.}~\bibnamefont {Dong}},\
  }\href {\doibase 10.3390/atoms3040474} {\bibfield  {journal} {\bibinfo
  {journal} {Atoms}\ }\textbf {\bibinfo {volume} {3}},\ \bibinfo {pages}
  {474} (\bibinfo {year} {2015})}\BibitemShut {NoStop}%
\bibitem [{\citenamefont {Gu}(2003)}]{gu2003a}%
  \BibitemOpen
  \bibfield  {author} {\bibinfo {author} {\bibfnamefont {M.~F.}\ \bibnamefont
  {Gu}},\ }\href {\doibase 10.1086/375135} {\bibfield  {journal} {\bibinfo
  {journal} {ApJ}\ }\textbf {\bibinfo {volume} {590}},\ \unskip\ \bibinfo
  {pages} {1131} (\bibinfo {year} {2003})}\BibitemShut {NoStop}%
\bibitem [{\citenamefont {Kwon}\ and\ \citenamefont {Lee}(2016)}]{kwon2016a}%
  \BibitemOpen
  \bibfield  {author} {\bibinfo {author} {\bibfnamefont {D.~H.}\ \bibnamefont
  {Kwon}}\ and\ \bibinfo {author} {\bibfnamefont {W.}~\bibnamefont {Lee}},\
  }\href {\doibase 10.1016/j.jqsrt.2015.11.009} {\bibfield  {journal} {\bibinfo
   {journal} {JQSRT}\ }\textbf {\bibinfo {volume} {170}},\ \unskip\ \bibinfo
  {pages} {182} (\bibinfo {year} {2016})}\BibitemShut {NoStop}%
\bibitem [{\citenamefont {Bar-Shalom}, \citenamefont {Klapisch},\ and\
  \citenamefont {Oreg}(2001)}]{barshalom2001a}%
  \BibitemOpen
  \bibfield  {author} {\bibinfo {author} {\bibfnamefont {A.}~\bibnamefont
  {Bar-Shalom}}, \bibinfo {author} {\bibfnamefont {M.}~\bibnamefont
  {Klapisch}}, \ and\ \bibinfo {author} {\bibfnamefont {J.}~\bibnamefont
  {Oreg}},\ }\href {\doibase http://dx.doi.org/10.1016/S0022-4073(01)00066-8}
  {\bibfield  {journal} {\bibinfo  {journal} {Journal of Quantitative
  Spectroscopy and Radiative Transfer}\ }\textbf {\bibinfo {volume} {71}},\
  \unskip\ \bibinfo {pages} {169} (\bibinfo {year} {2001})}\BibitemShut
  {NoStop}%
\bibitem [{\citenamefont {Peleg}\ \emph {et~al.}(1998)\citenamefont {Peleg},
  \citenamefont {Behar}, \citenamefont {Mandelbaum},\ and\ \citenamefont
  {Schwob}}]{peleg1998a}%
  \BibitemOpen
  \bibfield  {author} {\bibinfo {author} {\bibfnamefont {A.}~\bibnamefont
  {Peleg}}, \bibinfo {author} {\bibfnamefont {E.}~\bibnamefont {Behar}},
  \bibinfo {author} {\bibfnamefont {P.}~\bibnamefont {Mandelbaum}}, \ and\
  \bibinfo {author} {\bibfnamefont {J.~L.}\ \bibnamefont {Schwob}},\ }\href
  {\doibase 10.1103/PhysRevA.57.3493} {\bibfield  {journal} {\bibinfo
  {journal} {Phys. Rev. A}\ }\textbf {\bibinfo {volume} {57}},\ \unskip\
  \bibinfo {pages} {3493} (\bibinfo {year} {1998})}\BibitemShut {NoStop}%
\bibitem [{\citenamefont {Safronova}, \citenamefont {Safronova},\ and\
  \citenamefont {Beiersdorfer}(2009)}]{usafronova2009b}%
  \BibitemOpen
  \bibfield  {author} {\bibinfo {author} {\bibfnamefont {U.~I.}\ \bibnamefont
  {Safronova}}, \bibinfo {author} {\bibfnamefont {A.~S.}\ \bibnamefont
  {Safronova}}, \ and\ \bibinfo {author} {\bibfnamefont {P.}~\bibnamefont
  {Beiersdorfer}},\ }\href {\doibase
  http://dx.doi.org/10.1016/j.adt.2009.04.001} {\bibfield  {journal} {\bibinfo
  {journal} {Atomic Data and Nuclear Data Tables}\ }\textbf {\bibinfo {volume}
  {95}},\ \unskip\ \bibinfo {pages} {751} (\bibinfo {year}
  {2009})}\BibitemShut {NoStop}%
\bibitem [{\citenamefont {Behar}, \citenamefont {Mandelbaum},\ and\
  \citenamefont {Schwob}(1999)}]{behar1999a}%
  \BibitemOpen
  \bibfield  {author} {\bibinfo {author} {\bibfnamefont {E.}~\bibnamefont
  {Behar}}, \bibinfo {author} {\bibfnamefont {P.}~\bibnamefont {Mandelbaum}}, \
  and\ \bibinfo {author} {\bibfnamefont {J.~L.}\ \bibnamefont {Schwob}},\
  }\href {\doibase 10.1103/PhysRevA.59.2787} {\bibfield  {journal} {\bibinfo
  {journal} {Phys. Rev. A}\ }\textbf {\bibinfo {volume} {59}},\ \unskip\
  \bibinfo {pages} {2787} (\bibinfo {year} {1999})}\BibitemShut {NoStop}%
\bibitem [{\citenamefont {Preval}, \citenamefont {Badnell},\ and\ \citenamefont
  {O'Mullane}(2016)}]{preval2016a}%
  \BibitemOpen
  \bibfield  {author} {\bibinfo {author} {\bibfnamefont {S.~P.}\ \bibnamefont
  {Preval}}, \bibinfo {author} {\bibfnamefont {N.~R.}\ \bibnamefont {Badnell}},
  \ and\ \bibinfo {author} {\bibfnamefont {M.~G.}\ \bibnamefont {O'Mullane}},\
  }\href {\doibase 10.1103/PhysRevA.93.042703} {\bibfield  {journal} {\bibinfo
  {journal} {Phys. Rev. A}\ }\textbf {\bibinfo {volume} {93}},\ \unskip\
  \bibinfo {pages} {042703} (\bibinfo {year} {2016})}\BibitemShut {NoStop}%
\bibitem [{\citenamefont {Badnell}(1986)}]{badnell1986a}%
  \BibitemOpen
  \bibfield  {author} {\bibinfo {author} {\bibfnamefont {N.~R.}\ \bibnamefont
  {Badnell}},\ }\href {http://stacks.iop.org/0022-3700/19/i=22/a=023}
  {\bibfield  {journal} {\bibinfo  {journal} {Journal of Physics B: Atomic and
  Molecular Physics}\ }\textbf {\bibinfo {volume} {19}},\  \bibinfo {pages}
  {3827} (\bibinfo {year} {1986})}\BibitemShut {NoStop}%
\bibitem [{\citenamefont {Badnell}(1997)}]{badnell1997a}%
  \BibitemOpen
  \bibfield  {author} {\bibinfo {author} {\bibfnamefont {N.~R.}\ \bibnamefont
  {Badnell}},\ }\href {http://stacks.iop.org/0953-4075/30/i=1/a=005} {\bibfield
   {journal} {\bibinfo  {journal} {Journal of Physics B: Atomic, Molecular and
  Optical Physics}\ }\textbf {\bibinfo {volume} {30}}, ~\bibinfo {pages} {1}
  (\bibinfo {year} {1997})}\BibitemShut {NoStop}%
\bibitem [{\citenamefont {Badnell}(2011)}]{badnell2011a}%
  \BibitemOpen
  \bibfield  {author} {\bibinfo {author} {\bibfnamefont {N.~R.}\ \bibnamefont
  {Badnell}},\ }\href {\doibase http://dx.doi.org/10.1016/j.cpc.2011.03.023}
  {\bibfield  {journal} {\bibinfo  {journal} {Computer Physics Communications}\
  }\textbf {\bibinfo {volume} {182}},\ \bibinfo {pages} {1528} (\bibinfo
  {year} {2011})}\BibitemShut {NoStop}%
\bibitem [{\citenamefont {{Savin}}\ \emph {et~al.}(2002)\citenamefont
  {{Savin}}, \citenamefont {{Behar}}, \citenamefont {{Kahn}}, \citenamefont
  {{Gwinner}}, \citenamefont {{Saghiri}}, \citenamefont {{Schmitt}},
  \citenamefont {{Grieser}}, \citenamefont {{Repnow}}, \citenamefont
  {{Schwalm}}, \citenamefont {{Wolf}}, \citenamefont {{Bartsch}}, \citenamefont
  {{M{\"{u}}ller}}, \citenamefont {{Schippers}}, \citenamefont {{Badnell}},
  \citenamefont {{Chen}},\ and\ \citenamefont {{Gorczyca}}}]{savin2005a}%
  \BibitemOpen
  \bibfield  {author} {\bibinfo {author} {\bibfnamefont {D.~W.}\ \bibnamefont
  {{Savin}}}, \bibinfo {author} {\bibfnamefont {E.}~\bibnamefont {{Behar}}},
  \bibinfo {author} {\bibfnamefont {S.~M.}\ \bibnamefont {{Kahn}}}, \bibinfo
  {author} {\bibfnamefont {G.}~\bibnamefont {{Gwinner}}}, \bibinfo {author}
  {\bibfnamefont {A.~A.}\ \bibnamefont {{Saghiri}}}, \bibinfo {author}
  {\bibfnamefont {M.}~\bibnamefont {{Schmitt}}}, \bibinfo {author}
  {\bibfnamefont {M.}~\bibnamefont {{Grieser}}}, \bibinfo {author}
  {\bibfnamefont {R.}~\bibnamefont {{Repnow}}}, \bibinfo {author}
  {\bibfnamefont {D.}~\bibnamefont {{Schwalm}}}, \bibinfo {author}
  {\bibfnamefont {A.}~\bibnamefont {{Wolf}}}, \bibinfo {author} {\bibfnamefont
  {T.}~\bibnamefont {{Bartsch}}}, \bibinfo {author} {\bibfnamefont
  {A.}~\bibnamefont {{M{\"{u}}ller}}}, \bibinfo {author} {\bibfnamefont
  {S.}~\bibnamefont {{Schippers}}}, \bibinfo {author} {\bibfnamefont {N.~R.}\
  \bibnamefont {{Badnell}}}, \bibinfo {author} {\bibfnamefont {M.~H.}\
  \bibnamefont {{Chen}}}, \ and\ \bibinfo {author} {\bibfnamefont {T.~W.}\
  \bibnamefont {{Gorczyca}}},\ }\href {\doibase 10.1086/323388} {\bibfield
  {journal} {\bibinfo  {journal} {ApJS}\ }\textbf {\bibinfo {volume} {138}},\
  \unskip\ \bibinfo {pages} {337--370} (\bibinfo {year} {2002})}
  \BibitemShut {NoStop}%
\bibitem [{\citenamefont {Badnell}\ and\ \citenamefont
  {Pindzola}(1989)}]{badnell1989a}%
  \BibitemOpen
  \bibfield  {author} {\bibinfo {author} {\bibfnamefont {N.~R.}\ \bibnamefont
  {Badnell}}\ and\ \bibinfo {author} {\bibfnamefont {M.~S.}\ \bibnamefont
  {Pindzola}},\ }\href {\doibase 10.1103/PhysRevA.39.1685} {\bibfield
  {journal} {\bibinfo  {journal} {Phys. Rev. A}\ }\textbf {\bibinfo {volume}
  {39}},\ \unskip\ \bibinfo {pages} {1685} (\bibinfo {year}
  {1989})}\BibitemShut {NoStop}%
\end{thebibliography}

\newpage
\bibliographystyle{aipnum-cp}%

\end{document}